\documentclass[prb,preprint]{revtex4}%
\usepackage{graphicx}
\usepackage{amsmath}
\usepackage{ulem}
\usepackage{color}

\newcommand{\be}{\begin{equation}}
\newcommand{\ee}{\end{equation}}
\newcommand{\bea}{\begin{eqnarray*}}
\newcommand{\eea}{\end{eqnarray*}}
\newcommand{\bean}{\begin{eqnarray}}
\newcommand{\eean}{\end{eqnarray}}

\begin{document}

\draft
\title
{\bf Long distance coherent tunneling effect on the charge and heat
currents in serially coupled triple quantum dots}

\author{ David M T Kuo$^{1,2}$ and Yia-chung Chang$^{3,4}$}
\address{$^{1}$Department of Electrical Engineering and $^{2}$Department of Physics, National Central
University, Chungli, 320 Taiwan}


\address{$^{3}$Research Center for Applied Sciences, Academic Sinica,
Taipei, 11529 Taiwan}
\affiliation{$^4$ Department of Physics, National Cheng Kung University, Tainan, 701 Taiwan}

\date{\today}

\begin{abstract}
The effect of long distance coherent tunneling (LDCT) on the charge and heat currents
in serially coupled triple quantum dots
(TQDs) connected to electrodes is illustrated by using a combination of the extended
Hurbbard model and Anderson model. The charge and heat currents are calculated with a
closed-form Landauer expression for the transmission coefficient suitable for
the Coulomb blockade regime.  The physical parameters including bias-dependent quantum dot
energy levels, electron
Coulomb interactions, and electron hopping strengths are calculated
in the framework of effective mass theory for semiconductor TQDs. We demonstrate that the
effect of
LDCT on the charge and heat currents can be robust. In addition, it is shown that prominent
heat rectification behavior can exist in the TQD system with asymmetrical energy levels.
\end{abstract}

\maketitle
\section{Introduction}
Semiconductor quantum dots (QDs) have been advocated to be promising
candidates as qubits for the realization of solid state quantum
computer due to their long coherent time in charge and spin degrees
of freedom in comparison to their counterparts.$^{1,2}$ Many
experimental studies have been devoted to the coherent tunneling
behavior of serially coupled double quantum dots (DQDs).$^{3}$ The
serially coupled DQDs can be used as a spin filter when the Pauli
spin blockade condition is met.$^3$ To scale up quantum registers
based on QD arrays, one must have good control on the transport
properties of quantum dot arrays. It is expected that serially
coupled triple quantum dots (TQDs) can reveal the salient features
of the charge transport behavior in quantum dot arrays.$^{4-6}$ The
tunneling current spectra of serially coupled TQDs exhibit an
unexpected resonance structure arising from the long distance
coherent tunneling (LDCT) between the outer QDs.$^{4-6}$ The
interdot coupling strength decreases exponentially with the
separation between QDs. Therefore, the direct coupling strength
between the outer QDs of a TQD system, which are separated by a
large distance, is vanishingly small. However, it has been
demonstrated experimentally in Refs. 4 and 5 that the coherent
tunneling coupling between the outer QDs is not negligible due to
the middle QD-assisted tunneling, which can be understood from the
second-order perturbation theory.$^{7}$

Many theoretical studies have been devoted to the transport
properties of TQD systems. Topological effect on the electronic
properties of a TQD molecule was investigated in Ref. 8. The authors
of Ref. 9 studied the control of spin blockade by ac magnetic field
in TQDs. Weymann, Bulka and Barnas investigated the dark states in
transport through a triangle TQD.$^{10}$ The transport properties of
a serially coupled TQD have been studied by the master equation for
studying multiple electron spin blockade effect in Ref. 11. However
there still lacks a systematic analysis to illustrate the LDCT
effect on the tunneling current spectra of the serially coupled TQD
junction when the intradot and interdot Coulomb interactions are
included.

Besides the qubit aspect, the serially coupled QD arrays can also be
used as solid state coolers and power generators at nanoscale, which
is important in the integration of quantum device circuits.$^{12}$
The understanding of energy transfer and heat extraction of the QD
array is also crucial in the implementation of solid state quantum
register, because the heat accumulation will degrade the performance
of quantum computation. Unlike electronic nanodevices, it is still a
challenge to measure the heat transport in nanoscale
structures.$^{12}$ Up to date, most researches on the thermoelectric
properties of nanostructures connected to electrodes have been
restricted to theoretical studies.$^{13-28}$ References 13-18
investigated the optimization of figure of merit of QD junction
system in the linear response regime. The nonlinear thermoelectric
properties of nanostructures including QDs, molecules and the other
mesoscopic conductors can lead to attractive applications such as
thermal rectifiers, heat engines and thermal spintronics.$^{19-28}$
In this paper, the effect of LDCT on the charge and heat currents of
TQD junction system is revealed and analyzed in the presence of
intradot and interdot electron Coulomb interactions.

\section{Formalism}
Here we consider nanoscale semiconductor QDs, in which the energy
level separations are much larger than their on-site Coulomb
interactions and thermal energies. Thus, only one energy level for
each quantum dot needs to be considered. An extended Hubbard model
and Anderson model are employed to simulate a TQD connected to
electrodes.  The Hamiltonian of the TQD junction is given
by $H=H_0+H_{QD}$:

\begin{eqnarray}
H_0& = &\sum_{k,\sigma} \epsilon_k
a^{\dagger}_{k,\sigma}a_{k,\sigma}+ \sum_{k,\sigma} \epsilon_k
b^{\dagger}_{k,\sigma}b_{k,\sigma}\\ \nonumber &+&\sum_{k,\sigma}
V_{k,1}d^{\dagger}_{1,\sigma}a_{k,\sigma}
+\sum_{k,\sigma}V_{k,3}d^{\dagger}_{3,\sigma}b_{k,\sigma}+c.c
\end{eqnarray}
where the first two terms describe the free electron gas of left and
right electrodes. $a^{\dagger}_{k,\sigma}$
($b^{\dagger}_{k,\sigma}$) creates  an electron of momentum $k$ and
spin $\sigma$ with energy $\epsilon_k$ in the left (right)
electrode. $V_{k,\ell}$ ($\ell=1,3$) describes the coupling between
the electrodes and the first ($3$-th) QD.
$d^{\dagger}_{\ell,\sigma}$ ($d_{\ell,\sigma}$) creates (destroys)
an electron in the $\ell$-th dot.

\begin{small}
\begin{eqnarray}
H_{QD}&=& \sum_{\ell,\sigma} E_{\ell} n_{\ell,\sigma}+
\sum_{\ell} U_{\ell} n_{\ell,\sigma} n_{\ell,\bar\sigma}\\
\nonumber &+&\frac{1}{2}\sum_{\ell,j,\sigma,\sigma'}
U_{\ell,j}n_{\ell,\sigma}n_{j,\sigma'}
+\sum_{\ell,j,\sigma}t_{\ell,j} d^{\dagger}_{\ell,\sigma}
d_{j,\sigma},
\end{eqnarray}
\end{small}
where { $E_{\ell}$} is the spin-independent QD energy level, and
$n_{\ell,\sigma}=d^{\dagger}_{\ell,\sigma}d_{\ell,\sigma}$.
Notations $U_{\ell}$ and $U_{\ell,j}$ describe the intradot and
interdot Coulomb interactions, respectively. $t_{\ell,j}$ describes
the electron interdot hopping. Here we assume that the
interdot hopping and interdot Coulomb interaction in Eq. (2) are appreciable only between the nearest neighbor
QDs.

Using the Keldysh-Green's function technique$^{29}$, the charge
current of a TQD junction is calculated according to
\begin{equation}
J=\frac{2e}{h}\int d\epsilon {\cal T}(\epsilon)
[f_L(\epsilon)-f_R(\epsilon)].
\end{equation}
Meanwhile, the heat current that flows into  the right (left)
electrode from the TQD system is given by
\begin{equation}
Q_{R(L)}=\pm \frac{2}{h}\int d\epsilon {\cal T}(\epsilon)
(\epsilon-\mu_{R(L)})[f_L(\epsilon)-f_R(\epsilon)].
\end{equation}
We note that $Q_R+Q_L=Q_{Joule}=J\Delta V_a$, which indicates that
the heat flux dissipated from the TQD is equal to the electrical
power generated by Joule heating.$^{23,30}$ Note that in the open
circuit condition ($J=0$), which is the case for our consideration
in the study of electron heat rectification, we have
$Q_{Joule}=J\Delta V_a =0$ and $Q_R=-Q_L$. In Eqs.~(3) and (4),
${\cal T}(\epsilon)\equiv ({\cal T}_{1,3}(\epsilon) +{\cal
T}_{3,1}(\epsilon))/2$ is the transmission coefficient. ${\cal
T}_{\ell,j}(\epsilon)$ denotes the transmission function, which can
be calculated by using the on-site retarded and lesser Green's
functions. The transmission function in the weak interdot hopping
limit ($t_{\ell,j} \ll U_{\ell}$) has the following form,
\begin{equation}
{\cal T}_{\ell,j}(\epsilon)=-2\sum^{32}_{m=1}\frac{\Gamma_{\ell}(\epsilon)
\Gamma^{m}_{\ell,j}(\epsilon)}{\Gamma_{\ell}(\epsilon)+\Gamma^{m}_{\ell,j}(\epsilon)}
\mbox{Im}G^r_{\ell,m}(\epsilon),
\end{equation}
where "Im" means taking the imaginary part of the function that
follows, and
\begin{equation}
G^r_{\ell,m}(\epsilon)=p_{\ell,m}/(\mu_{\ell}-\Pi_{\ell,m}-\Sigma^m_{\ell,j}),
\end{equation}
where $\mu_{\ell}=\epsilon-E_{\ell}+i\Gamma_{\ell}/2$. Note that
$\Gamma_{\ell}=0$ when $\ell \neq 1,3$. $\Gamma_{\ell}$ denotes the
tunneling rate for electron tunneling from QD $\ell$ to the
electrode. In Eqs. (5) and (6), we have $\ell \neq j$.
$\Pi_{1(3),m}$ denotes the sum of Coulomb energies between one
electron in the first (third) QD and other electrons present in its
neighboring QDs in configuration $m$, and
$\Gamma^m_{\ell,j}(\epsilon)=-2Im\Sigma^m_{\ell,j}(\epsilon)$, where
$\Sigma^m_{\ell,j}$ denotes the self energy resulting from electron
hopping from QD $\ell$ to QD $j$ through channel $m$. The detailed
expressions of probability weight $p_{\ell,m}$ as well as
$\Pi_{\ell,m}$ and $\Sigma^m_{\ell,j}$ can be found in Ref. 31.

The factor 2 in Eqs. (3) and (4) comes from electron spin
degeneracy.
$f_{L(R)}(\epsilon)=1/[e^{(\epsilon-\mu_{L(R)})/k_BT_{L(R)}}+1]$
denotes the Fermi distribution function for the left (right)
electrode. $\mu_L$ and $\mu_R$ are the chemical potentials of the
left and right leads, respectively, with their average denoted by
$E_F=(\mu_L+\mu_R)/2$. $\Delta V_a=(\mu_L-\mu_R)/e$ is the voltage
across the TQD junction. $T_{L(R)}$ is the equilibrium temperature
of the left (right) electrode.  $e$ and $h$ denote the electron
charge and Planck's constant, respectively.

To study the LDCT effect on the charge and heat currents of Eqs. (3)
and (4), it is important to provide reasonable physical parameters.
So far, the exact solution of ${\cal T}(\epsilon)$, which are valid
for strong-coupling regime (with $t_{\ell,j}$ comparable to
$U_{\ell,j}$), has not been reported due to the many body
effect.$^{8,11}$  Although the first principles method is often used
to  calculate ${\cal T}(\epsilon)$ directly, it can not handle the
charge and heat currents in the Coulomb blockade regime since it is
a mean-field approach.$^{25}$ Here, we use the extended
Hubbard-Anderson model ($H=H_0+H_{QD}$) to illustrate the transport
and thermoelectric properties of three disk-like (or cone-shaped)
GaAs QDs embedded in Al$_x$Ga$_{1-x}$As connected to electrodes. The
physical parameters for $U_{\ell,j}$ and $t_{\ell,j}$ used in
$H_{QD}$ can be calculated in the framework of effective mass
method.$^{32}$  The effective-mass equation for a coupled QD (CQD)
system is given by \be [-\nabla \frac {\hbar^2} {2m^*(\rho,z)}
\nabla + V_{CQD}(\rho,z)+V_{sc}(\textbf{r})] \psi ({\bf r}) =  E
\psi({\bf r}), \ee where $m^{*}_e(\rho,z)$ denotes the
position-dependent electron effective mass. We adopt effective
masses $m^*_{GaAs}=0.063~m_e$ for GaAs and $m^*=0.096~m_e$ for
$Al_{0.4}Ga_{0.6}As$. $V_{CQD}(\rho,z)$ is approximated by a
constant potential $V_0=-0.2496~eV$ in the QD region and zero in the
barrier region. Its value is determined by the conduction band
offset between GaAs and $Al_{0.4}Ga_{0.6}As$.
$V_{sc}(\textbf{r})=\frac{e^2}{\varepsilon_0} \int d\textbf{r}'
n_e(\textbf{r}')/|\textbf{r}-\textbf{r}'|$ denotes the
self-consistent potential caused by the electrostatic interaction
with the charge density associated with the other particles in the
system. Note that we consider the position-independent dielectric
constant $\varepsilon_0=15.5$. For the purpose of constructing
approximate wave functions, we place the system in a large
cylindrical confining box with the length $L $ and radius $R$ ($L$
and $R $ must be much larger than those of CQD). Here we choose $L =
60 $~nm and $R = 40~nm$. We solve Eq. (7) by the Ritz variational
method.$^{32}$ The wave functions are expanded in a set of basis
functions, which are chosen to be products of Bessel functions of
$\rho$ and sine functions of $z$, \be
\psi_{n,\ell,m}(\rho,z,\phi)=J_{\ell}(\beta_n\rho)e^{i\ell \phi}
\sin(k_m(z+L/2)), \ee where $k_m=m\pi/L,m=1,2,3..$, $J_{\ell}$ is
the Bessel function of  order $\ell$ and $\beta_n R$ is the $n$th
zero of $J_{\ell}$. The expression of the matrix elements of Eq. (7)
can be readily obtained. Forty-five sine functions multiplied by
fifteen Bessel functions for each angular function ($\ell$ = 0 or 1)
are used to diagonalize $H_{CQD}$. A convergence check (by
increasing the basis functions) indicates that the lowest two
confined states are accurate to within $0.1~meV$ with the current
set of bases.

Figure~1 shows (a) energy levels  and (b) electron hopping strengths
for two different shapes of GaAs QDs (cone and disk) as functions of
the gap distance ($D$) between two QDs. The height and radius of
each QD are $L_1=L_2=5~nm$ and $R_1=R_2=10~nm$. Let $t_c$ denotes
the hopping strength between adjacent QDs in a tight-binding model.
The energy separation between the bonding ($E_{BD}=E_0+t_c$)and
antibonding ($E_{AB}=E_0-t_c$) states increases with decreasing gap
distance ($D$). The electron hopping strength ($t_{\ell,j}=t_c$) is
smaller than $0.1 ~meV$ when the gap distance $D$ is larger than
$10~nm$. When $D=4~nm$, $t_c$ is approximately $3.5~meV$. Note that
$t_{\ell,j}$ as a function of gap distance can be fitted very well
by an exponential decay function, which can be used to estimate the
very week coupling for two QDs separated far away. As seen in
Fig.~1(b), when the gap distance is doubled, $t_{\ell,j}$ reduces to
{~$0.36~meV$ (a factor 10 smaller)}. Therefore, it is adequate to
keep the inter-dot coupling $t_{\ell,j}$ only for adjacent dots,
since $t_{1,3}$ (between two outer dots) is significantly smaller.
To evaluate the electron Coulomb interaction strengths, we calculate
the intradot and interdot Coulomb interactions as functions of the
gap distance ($D$) for two disk-shaped QDs: dot A with $L_1=5~nm$
and $R_1=10~nm$ and dot B with $L_2=5.5~nm$ and $R_2=11~nm$. The
results are shown in Fig.~1(c). The position dependence for intradot
Coulomb interactions are noticeable only  for small $D$, where the
leak-out of QD wave function is appreciable.

In the presence of an applied bias, the resulting electric field
leads to the energy level shift. To examine this effect, we added a
term $-eF(z-z_0)$ in Eq.~(7). $F$ and $z_0$ denote the
electric-field strength and the middle point of the junction,
respectively. Figure~2 shows the lowest two energy levels as
functions of electric-field strength ($F$) for disk-shaped QDs with
(a) identical QD sizes ($R_1=R_2=10~nm$)and (b) different QD sizes
($R_1=10~nm$ and $R_2=9~nm$), while $L_1=L_2=5~nm$. In Fig.~2(a),
the energy gap (~$2t_c$) arises from the resonant tunneling between
$E_1$ and $E_2$ levels in the absence of $F$. When $F$ increases,
this resonant coupling is diminished (off resonance) due to the
increased separation of $E_1$ and $E_2$, which is approximately
linear in $F$. In Fig.~2(b), $E_1$ and $E_2$ levels are "off
resonance" in the absence of $F$. However, $F$ can be tuned to bring
the $E_1$ and $E_2$ levels in resonance [see Fig.~2(b) near $F = 3
kV/cm$]. The results of Fig. 2 indicate that the energy level shift
as well as $t_{\ell,j}$ and $U_{\ell}(U_{\ell,j})$ will
significantly affect the behaviors of charge and heat currents.

For electrodes made of heavy-doped semiconductors, the Fermi energy
of semiconductor electrode depends on the carrier concentration. For
example, the carrier concentration $n=2\times 10^{18}/cm^3$ in the
$GaAs$ electrodes leads to $E_F=91.76~meV$. Therefore, we have
$E_{\ell}-E_F=7.24~meV$ for disk-shaped QDs described by the solid
lines in Fig.~1(a). The level $E_{\ell}-E_F$ can be tuned by a gate
electrode or the carrier concentration of electrodes. In the
following numerical calculations, we consider a GaAs/AlGaAs TQD
junction with gap distance $D=8nm$ and and QD size $L_{\ell}=5~nm$.
Using the effective-mass model described above, we obtain
$U_{\ell}=200\Gamma_0$, $U_{\ell,j}=66\Gamma_0$, and
$t_{\ell,j}=3.6\Gamma_0$, where $\Gamma_0=0.1~meV$. This set of
physical parameters satisfies the condition of $U_{\ell}
> U_{\ell,j} \gg t_{\ell,j}$ and $U_{\ell} \gg \Gamma_{\ell}$,
which are required in keeping the validity of Eq.~(5) for ${\cal
T}(\epsilon)$.$^{31}$ The tunneling rates $\Gamma_{\ell}$ can also
be accurately determined by a stabilization method.$^{32}$

\section{Results and discussion}

\subsection{LDCT for charge current}
The results of Fig.~2 show that the bias-dependent shift of energy
level ($E_{\ell}$) in each dot can be approximately determined
according to the expression $\epsilon_{\ell} = E_{\ell} +
\eta_{\ell}e\Delta V_{a}$, where $\eta_{\ell}$ denotes the fraction
of voltage difference shared by QD $\ell$. The value of
$\eta_{\ell}$ depends on the location, shape and dielectric constant
of the QD. When the dielectric constants of the QD and the
surrounding material are similar, the voltage difference is almost
uniformly distributed among QDs. Let $d_{\ell}$ denotes the center
position of QD $\ell$ with respect to the mid point of the junction
and the separation of two electrodes is $D_{LR}$, then the
electrostatic potential energy due to the uniform electric field
seen by an electron in QD ${\ell}$ is simply $V({\bf r}-d_{\ell}\hat
z)=[d_{\ell}+(z-d_{\ell})](-e\Delta V_{a}/D_{LR})$ ($z$ is along the
direction of transport). For weak field and symmetric wave function
in each QD, the linear $(z-d_{\ell})$ term vanishes, and the energy
correction due to second-order Stark effect is insignificant. Thus,
we have $\eta_{\ell}=d_{\ell}/D_{LR}$. For the TQD junction, we
assume $d_1=-d_3$ and $\eta_1=-\eta_3$.

We first calculate the tunneling current for
$\Gamma_L=\Gamma_R\equiv \Gamma$ (with $\Gamma=0.3\Gamma_0$) under
thermal equilibrium.  Figure~3(a) shows the tunneling current as a
function of {the applied bias} $\Delta V_a$ for a GaAs/AlGaAs TQD
junction with staircase energy levels ($E_{1}=E_F+9\Gamma_0$,
$E_2=E_F+6\Gamma_0$ and $E_3=E_F+3\Gamma_0$), $\eta_1=-\eta_3=0.24$,
and $D_{LR}=54~nm$. Such an energy level arrangement was also
considered in Ref.~6. We noticed a negative differential conductance
(NDC) behavior. This is an essential feature for resonant tunneling
junction due to off-resonance process.$^{4-6}$ There is an
unexpected resonance peak at $\Delta V_a=125\Gamma_0$ labeled by
$p_{3}$, whose contribution is mainly from the 3rd configuration as
described in Ref.~31. This $p_3$ resonance peak can be suppressed by
decreasing temperature. The structure labeled by $p_3$ arises from
the LDCT between the outer QDs associated with charging effect under
the resonant condition with energy levels $\epsilon_1\equiv
E_1+0.24e\Delta V_a =E_3+U_{23}-0.24e\Delta V_a \equiv \epsilon_3$.
Note that $\epsilon_2\equiv E_2+U_{23}$ and $E_2$ do not have to be
resonant with $\epsilon_1=\epsilon_3$, a main feature of LDCT. The
$p_3$ structure indicates that the middle QD can assist the electron
tunneling between outer QDs separated by a large distance through a
new channel in the presence of electron Coulomb interaction. In the
$p_3$ configuration, there is one electron in level $E_3$ with the
same spin $\sigma$ as the electron entering the junction from the
electrode [see the inset of Fig.~3(a)]. Lowering the temperature
decreases the probability of electron occupation in level $E_3$
($N_3$). Therefore, the peak of $p_3$ decreases with decreasing
temperature. It is noted that even at extremely low temperature, a
residue value of $N_3$ still exists. The tiny structure labeled by
$p_3$ at very low temperature can be resolved in the differential
conductance.

In order to illustrate the effect of electron Coulomb interactions,
we also show in Fig.~3(a) the tunneling current at $k_BT=6\Gamma_0$
(curve with triangle marks) for the case without Coulomb
interactions (i.e. $U_{\ell}=U_{\ell,j}=0$). In this case, the
structure labeled by $p_3$ vanishes, whereas the magnitude of $J$ is
enhanced.  We note that Fig. 3(a) displays a nearly
temperature-independent thermal broadening effect, which is very
different from that of a single QD junction.$^{13,14}$ Such a
"nonthermal" broadening effect is a common feature for serially
double QD junction system.$^{15,33}$ Reference 33 pointed out that
such a "nonthermal" broadening effect allows serially double QDs to
function as low-temperature filters. The detailed description of
low-temperature filter can be found in Ref. 33, where the electron
Coulomb interactions were neglected.

Because it is difficult to get an analytic expression of tunneling
current shown in Fig.~3(a), we  illustrate the effect of LDCT on the
electrical conductance of the TQD junction, where the QD levels are
aligned with $E_F$. For this case, simple expressions can be used to
reveal the LDCT effect. We show in Fig.~3(b) the electrical
conductance ($G_e$) as a function of gate voltage $V_g$, which is is
applied to tune the middle-dot level, $E_2$. Note that \be G_e
=\frac {2e^2}h \int d\epsilon {\cal T}(\epsilon) (-\frac {\partial
f}{\partial \epsilon}). \ee Here,  $E_1=E_3=E_F$ and
$E_2+eV_g=E_F+eV_g$. The trend of $G_e$ with respect to $eV_g$ can
be well explained by Eq.~(3) with transmission coefficient ${\cal
T}(\epsilon)\approx {\cal T}^1(\epsilon)=({\cal
T}^1_{1,3}(\epsilon)+{\cal T}^1_{3,1}(\epsilon))/2$.
\begin{equation}
{\cal T}^1(\epsilon)= \frac{\Gamma_L\Gamma_R t^2_{eff}(\epsilon)
p_1} {|\mu_1\mu_3-t_{eff}(\epsilon)
~\mu_3-t_{eff}(\epsilon)~\mu_1|^2},
\end{equation}
where $\mu_1=\epsilon-E_1+i\Gamma/2$ and
$\mu_3=\epsilon-E_3+i\Gamma/2$.
$t_{eff}(\epsilon)=t^2_c/(\epsilon-E_2-eV_g)$.

In the absence of $U_{\ell}$ and $U_{\ell,j}$, a similar expression
to Eq.~(10) can also be found in Ref.~34. The authors of Ref.~34
considered the effect of electron Coulomb interactions within the
Hartree-Fock approximation. With this approximation, the electron
occupation numbers will appear in the denominator of Eq.~(10). This
will lead to a fractional charge picture. In our procedure used for
calculating the retarded and lesser Green functions (which is beyond
the Hartree-Fock approximation), the electron occupation numbers and
two-particle correlation functions appear only in the probability
weight of each configuration.$^{31,35}$ The picture with integer
charges appearing in the denominators of Eq. (6) is consistent with
that of the master equation method.$^{11}$ In this approach, we only
considered the one-particle occupation number and on-site two
particle correlation functions in the probability weights. For weak
interdot hopping strength ($t_{\ell,j}/\Gamma < 1$), the
approximation which neglects the two-particle interdot correlation
functions and higher order functions can get results quite close to
those obtained by solving the equation of motion exactly (i.e.
including all correlation functions) via numerical computation,
which has been done and will be reported elsewhere.$^{36}$ Although
such an approximation is not so accurate for $t_{\ell,j}/\Gamma > 1$
when QD energy levels are degenerate, the procedure considered in
Eq. (5)$^{16,31}$ is justified as long as $|E_{\ell}-E_j|/t_{\ell,j}
\gg 1$.

 We note that ${\cal T}(\epsilon)$ is equal to ${\cal
T}^1(\epsilon)$ with $p_1=1$ in the absence of electron Coulomb
interactions.$^{31}$ With finite electron Coulomb interactions we
have $p_1=(1-N_1)(1-2N_2+c_2)(1-2N_3+c_3)$, where $N_{\ell}$ is the
one-electron occupancy in $E_\ell$ level. Note that $N_1=N_3$ for
the symmetrical configuration shown in the inset of Fig.~3(b). The
probability of two-electron occupancy $c_{\ell}$ in each QD can be
assumed zero due to the large value of $U_{\ell}$. The curve with
triangle marks in Fig.~3(b) shows $G_e$ in the absence of electron
Coulomb interactions at $k_BT=3\Gamma_0$. We find that $G_e$
increases initially, reaching a maximum, then decreases as $V_g$
increases. When the electron Coulomb interactions are included, the
probability factor $p_1$ becomes $V_g$ dependent (as shown in the
curve with squares), which modifies the behavior of $G_e$ as shown
in the solid, dashed, and dotted curves at $k_BT=1, 2$, and
$3\Gamma_0$.

Based on Eq.~(10), the solution of $G_e$ might be expressed in terms
of the poly-Gamma functions.$^{33}$ Rather than using the complicate
poly-Gamma functions, we derive some simple expressions from
Eq.~(10) in suitable limits to gain better understanding of the
behavior of $G_e$.  In the linear response regime with respect to
$\Delta V_a$, the energy levels shifted by $\Delta V_a$ can be
neglected in Eq.~(10). Since $E_1=E_3=E_F$ here, Eq.~(10) reduces to
${\cal T}^1(\epsilon)= \frac{p_1\Gamma_L\Gamma_R t_c^4}
{|\mu_1(\mu_1(\epsilon-E_2-eV_g)-2t^2_{c})|^2}$. The denominator can
be rewritten in the form $|(\epsilon-E_F+i\Gamma/2)(\epsilon-\tilde
\epsilon_+) (\epsilon-\tilde \epsilon_-)|^2 $, where \be \tilde
\epsilon_\pm=E_F-i\Gamma/2+[(eV_g+i\Gamma/2) \pm
\sqrt{(eV_g+i\Gamma/2)^2+8t_c^2}]/2 \ee  denote the energy positions
of two poles in addition to the pole at $E_F-i\Gamma/2$. At $V_g=0$,
the two poles $\tilde \epsilon_\pm= E_F-i\Gamma/4\pm \sqrt{2}t_c$
when $t_c \gg \Gamma$, which are located far from $E_F$ (since
$t_c\gg \Gamma$) and their contributions to $G_e$ become negligible.
Thus, at $V_g=0$ we have \be G_e \approx \frac{e^2}{h} \frac{p_1\pi
\Gamma}{2k_BT}, \ee which is dominated by the pole at
$\epsilon=E_F$.However, when $V_g$ increases, the two poles at
$\tilde \epsilon_\pm$ move up in energy [See Eq.~(11)] with the
lower pole approaching the level $E_F$, which gives appreciable
contribution to $G_e$ when the pole is in the range covered by the
function $(-\frac {\partial f}{\partial
\epsilon})=1/[4k_BT\cosh^2(\frac {\epsilon-E_F} {2k_BT})]$ appearing
in Eq.~(9). This explains the initial rise of $G_e$ (for $p_1=1$)
with respect to $V_g$ at finite temperatures as seen in Fig.~3(b).
As  temperature approaches zero, the function $(-\frac {\partial
f}{\partial \epsilon})$ turns into a delta function with
$\epsilon=E_F$, then we have $G_e =\frac {2e^2} h {\cal
T}(E_F)\approx (2e^2/h)p_1/[1+(eV_g\Gamma/(4t_c^2))^2]$. Thus, at
$T=0$, the maximum of $G_e$ would occur at $V_g=0$.

To understand the decrease of $G_e$ for large values of $V_g$ we
consider another asymptotic expression (for $k_BT > \Gamma > t_{eff}$)
\begin{equation}
G_e\approx \frac{e^2}{h} \frac{\pi \Gamma}{2k_BT} \frac{p_1
t^2_{eff}}{t^2_{eff}+\Gamma^2/4},
\end{equation}
where $t_{eff}=t^2_c/(eV_g)$. As $V_g$ approaches infinity, we
obtain an insulating state ($G_e \rightarrow 0$).

The curves with circle marks are calculated by Eq.~(13) for the case
including Coulomb interactions at $k_BT=1\Gamma_0$ and excluding
Coulomb interactions at $k_BT=3\Gamma_0$ (i.e. $p_1=1$). We find
good agreement between results obtained by Eq.~(10) and by the
asymptotic expression (13) for $eV_g > 43 \Gamma_0$. The factor
$k_BT$ appearing in the denominator of Eqs.~(12) and (13) also
explains why $G_e$ is suppressed with increasing temperature, as
seen in Fig.~3(b). The simple expression of Eq.~(13) is the
manifestation of the result for a DQD with effective coupling
strength $t_{eff}$.$^{33}$ It is convenient to use Eq.~(13) to
illustrates the effect of LDCT between the outer QDs separated by a
large distance.

For $t_c=0.36~meV$ (with $D=8~nm$ between the middle QD and the outer
QD), we obtain $t_{eff}=25.9~\mu eV$ for $eV_g=5~meV$ between outer
QDs separated by barriers with total thickness $D=16~nm$. (Note: the
width of the middle QD does not count toward the gap distance. Only
the barrier thickness counts) For such a gap distance, the direct
coupling for outer GaAs/GaAlAs QDs is $t_{13}=1.73 \mu eV$, which
is negligible compared to $t_{eff}$. The effect of LDCT is very
useful for improving the entanglement between qubits stored in
distant QDs.$^{4,5}$ Next, we investigate how LDCT influences the
electron heat rectification of TQD junctions.

\subsection{LDCT for heat current}

To study the direction-dependent heat current, we let $T_L = T_0 +
\Delta T/2$ and $T_R = T_0-\Delta T/2$, where $T_0 = (T_L + T_R)/2$
is the average of equilibrium temperatures of two side electrodes
and $\Delta T = T_L -T_R $ is the temperature difference across the
junction. We have numerically solved Eqs.~(3) and (4) for TQD
junctions. We first determine the nonlinear Seebeck coefficient
$S=e\Delta V_{th}/k_B\Delta T$ (thermal voltage yielded by $\Delta
T$) by solving Eq.~(3) with J = 0 (the open circuit condition) for a
given $\Delta T$, $T_0$ and an initial guess of the average
one-particle and two-particle occupancy numbers, $N$ and $c$ for
each QD, which can be found in Ref. 31. Due to $J=0$, we have
$Q_{Joule}=0$. Once $\Delta V_{th}$ is solved, we then use Eq. (4)
to compute the heat current. The nonlinear Seebeck coefficient of a
single molecule was studied in references [25,26] for the
applications of thermal spintronics. Fig. 4(a) shows the electron
heat current ($Q=Q_R$) as a function of temperature bias $\Delta T$
for various values of energy alignment $\Delta_{F}=E_3-E_F$ (while
keeping $E_1=E_3+2\Delta_0$, and $E_2=E_3+\Delta_0$) at
$T_0=26\Gamma_0$, and $\Gamma_L=\Gamma_R=7\Gamma_0$. Note that
$D_{LR}=43~nm$, we have $\eta_1=-\eta_3=0.31$. The energy levels of
the TQD have a staircase structure with step height
$\Delta_0=20\Gamma_0$. We considered $\Delta_0=20\Gamma_0$, which is
larger than that considered in Fig. 3, for observing electron heat
rectification in a wide temperature range. The results of Fig.~4(a)
indicate an asymmetrical heat current (rectification effect) which
depends on $\Delta_{F}$. When $T_0$ is larger than $\Delta_{F}$, the
rectification effect is seriously suppressed. The dash-dotted line
is almost symmetric, which means $Q$ is linearly proportional to
$\Delta T$.

To further enhance the asymmetrical behavior, we study the heat
current ($Q$) for various values of  $T_0$ at
$\Delta_{F}=40\Gamma_0$ as shown in Fig.~4(b). Other physical
parameters are the same as those for the solid line in Fig.~4(a).
When the averaged temperature $T_0$ goes down, the forward heat
current ($Q_F$) increases in the forward temperature bias ($\Delta T
> 0$), whereas the backward heat current ($Q_B$) decreases in the
reversed temperature bias ($\Delta T < 0$). The asymmetrical $Q$
behavior with respect to $\Delta T$ is enhanced with decreasing
$T_0$. Because QD energy levels are shifted by the thermal voltage
($\Delta V_{th}$), we show in Fig.~4(c) and 4(d) the thermal voltage
$\Delta V_{th}$ yielded by $\Delta T$, corresponding to Fig.~4(a)
and 4(b), respectively.  As shown in Fig.~4(c), when
$\Delta_F=10\Gamma_0$ the thermal voltage $\Delta V_{th}$ produced
is rather small, which is insufficient to give rise to  noticeable
nonlinear heat behavior with respect to $\Delta T$ [See dash-dotted
line in Fig.~4(a)]. Figure~4(d) shows the increase of nonlinear
behavior in $Q$  due to the enhancement of $\Delta V_{th}$. In
addition, the results of Fig.~4 also indicate that there exists a
nonlinear relationship between $Q$ and $\Delta V_{th}$. The results
of Fig.~4(b) (large heat current in $\Delta T > 0$ and small heat
current in $\Delta T < 0$) can be understood by the following. The
transmission coefficient ${\cal T}^m_{\ell,j}(\epsilon)$ of the
dominant configuration is proportional to the joint density of
states (JDOS), which is the product of spectral functions arising
from three resonant poles {(see Eq.~(7) of Ref. 31)}. The thermal
voltage $\Delta V_{th}$ causes shift in the QD levels. Thus, for a
TQD system with $E_1>E_2>E_3$ under zero bias (see the inset of
Fig.~4(c)), a positive $\Delta T$ can bring the levels close to
resonance, while a negative  $\Delta T$ will cause them further
apart from resonance, resulting a increases (decreases) in JDOS when
$\Delta T > 0 $ ($\Delta T < 0$).

Next, we examine the LDCT effect on the rectification behavior of
the TQD junction. Figure~5 shows the heat current and thermal
voltage as functions of temperature bias for
$\Delta_F=E_3-E_F=40\Gamma_0$ and $E_1=E_3+25\Gamma_0$ for various
values of $E_2$. As $E_2$ increases from $E_3+12.5\Gamma_0$ to
$E_3+25\Gamma_0$, the thermal voltage $\Delta V_{th}$ increases. In
comparison to the results of Figs.~4(b), the heat current of forward
temperature bias in Fig.~5 is enhanced significantly, which is
attributed to the enhancement of JDOS resulting from the better
alignment of resonant poles. {For the forward temperature bias
($\Delta T=12.5\Gamma_0$), the energy levels of outer QDs can be
aligned at $\Delta V_{th}=-40.3\Gamma_0$ (LDCT resonant level
$E_{LDCT}=E_F+52.5\Gamma_0$), while the middle QD energy level is
misaligned with $E_{LDCT}$, which leads to an effective tunneling
coupling $t_{eff}=t^2_c/(E_2-E_{LDCT})$.} From the results of
Fig.~5, we see that the LDCT can also improve the heat rectification
behavior for two distant QDs.$^{28}$ To observe such an electron
heat rectification effect shown in Fig. 4(b) and 5, the magnitude of
phonon heat current $Q_{ph}$ must not be dominant over the electron
heat current. To reduce $Q_{ph}$, we can design a QD array in which
the barriers (AlGaAs) have a small cross section (see the inset of
Fig.~5(a)) to produce a phonon bottleneck effect.$^{28}$ Although
many studies have been devoted to the design of phonon or photon
heat rectifiers,$^{37-39}$ these designs are not compatible with the
fabrication technique of solid state quantum register circuit. So
far, few experiments have observed the heat rectification
effect.$^{40}$

\section{Summary}
We have theoretically studied the effect of LDCT on the charge and
heat currents of a TQD junction in the Coulomb blockade regime. In
the presence of intradot and interdot Coulomb interactions the
closed form Landauer expression for transmission coefficient
provides a useful analysis for clarifying the influence of electron
Coulomb interactions on the LDCT effect. The middle QD can mediate
the coherent tunneling between distant outer QDs.  An interesting
electron heat rectification effect of the TQD junction is
demonstrated by considering a staircase-like alignment of energy
levels. Using the nonlinear Seebeck effect ($e\Delta
V_{th}/k_B\Delta T$), we can control the electron resonant process
of the TQD junction by temperature bias to observe heat
rectification.



\begin{flushleft}
{\bf Acknowledgments}\\
\end{flushleft}
This work was supported in part by the National Science Council of
the Republic of China under Contract Nos. NSC
101-2112-M-008-014-MY2, and NSC 101-2112-M-001-024-MY3.

\mbox{}\\
E-mail address: mtkuo@ee.ncu.edu.tw\\
E-mail address: yiachang@gate.sinica.edu.tw\\


\newpage
\section*{Figure captions}
Fig. 1. (a)Energy levels  and (b) electron hopping strength of
identical double quantum dots (GaAs/AlGaAs) as a function of gap
distance ($D$). (c) Intradot and interdot Coulomb interactions as a
function of D for Dot A and dot B with, respectively, the height
$L_1=~5nm$ (radius $R_1=2L_1$) and $L_2=5.5~nm$ (radius $R_2=2L_2$).

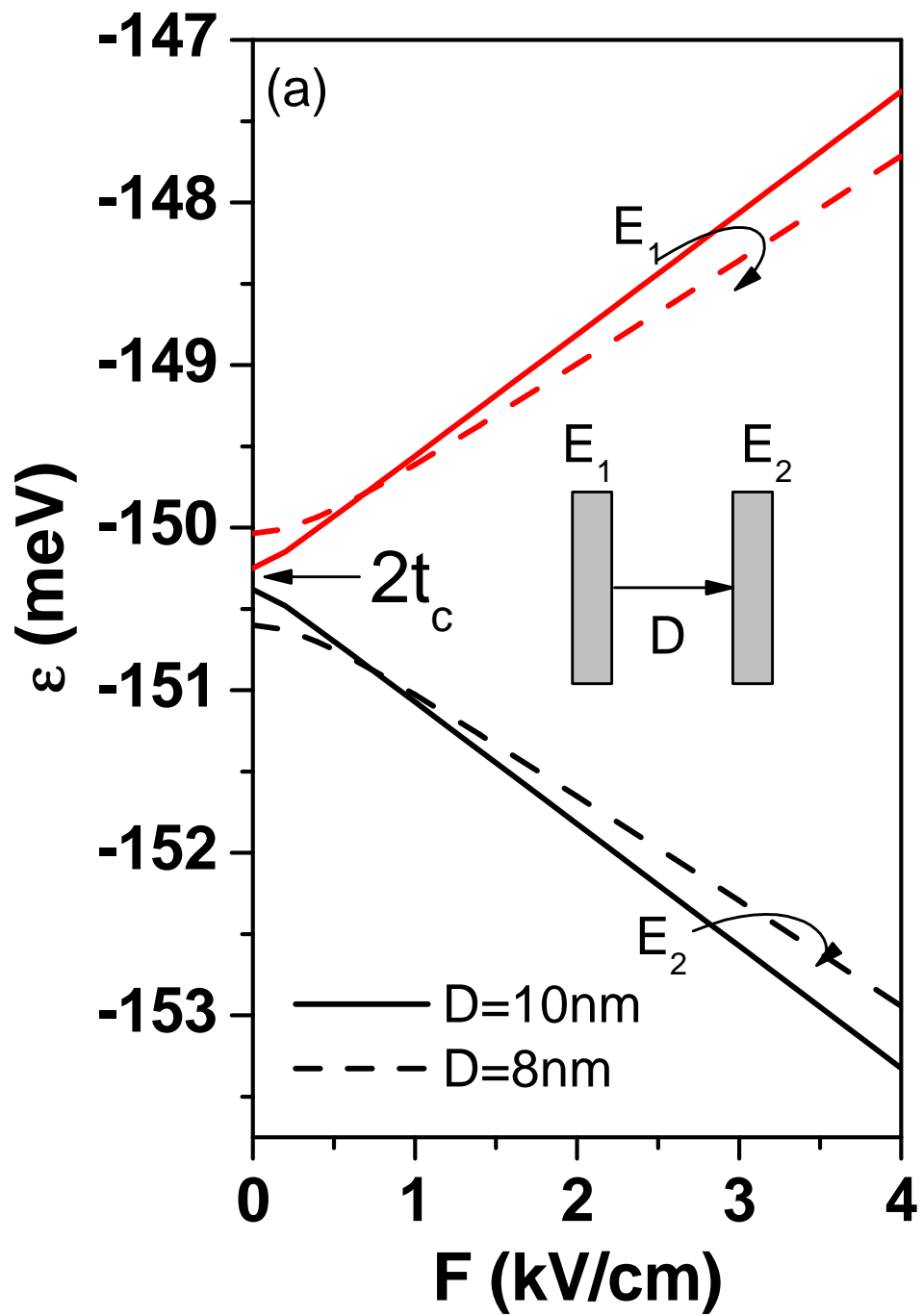
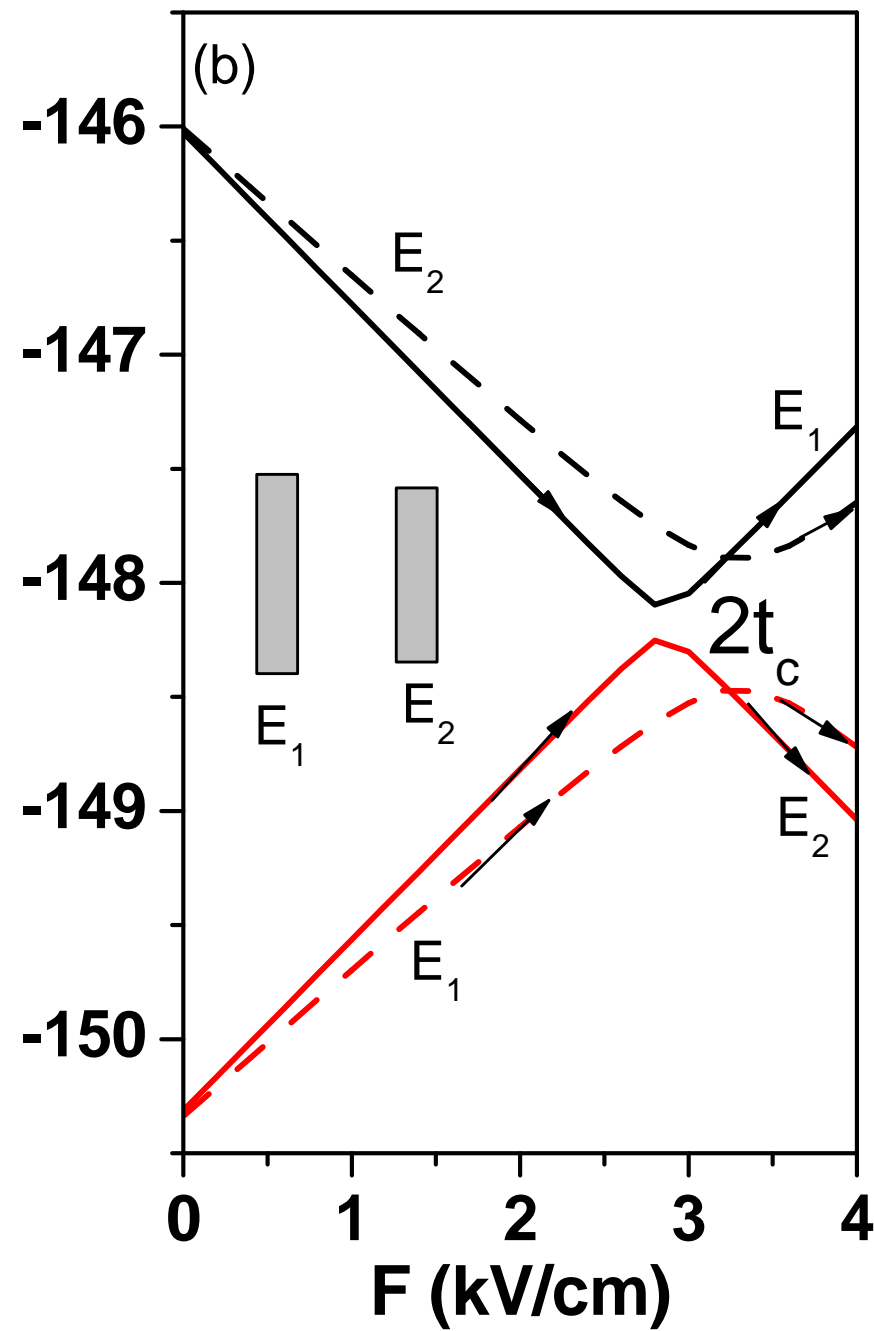
Fig. 2. The lowest two energy levels of coupled QDs as a function of
electric field strength ($F$) for two different gap distances. (a)
Identical QD sizes, and (b) different QD sizes. The other physical
parameters are the same as those for solid lines of Fig.~1(a).

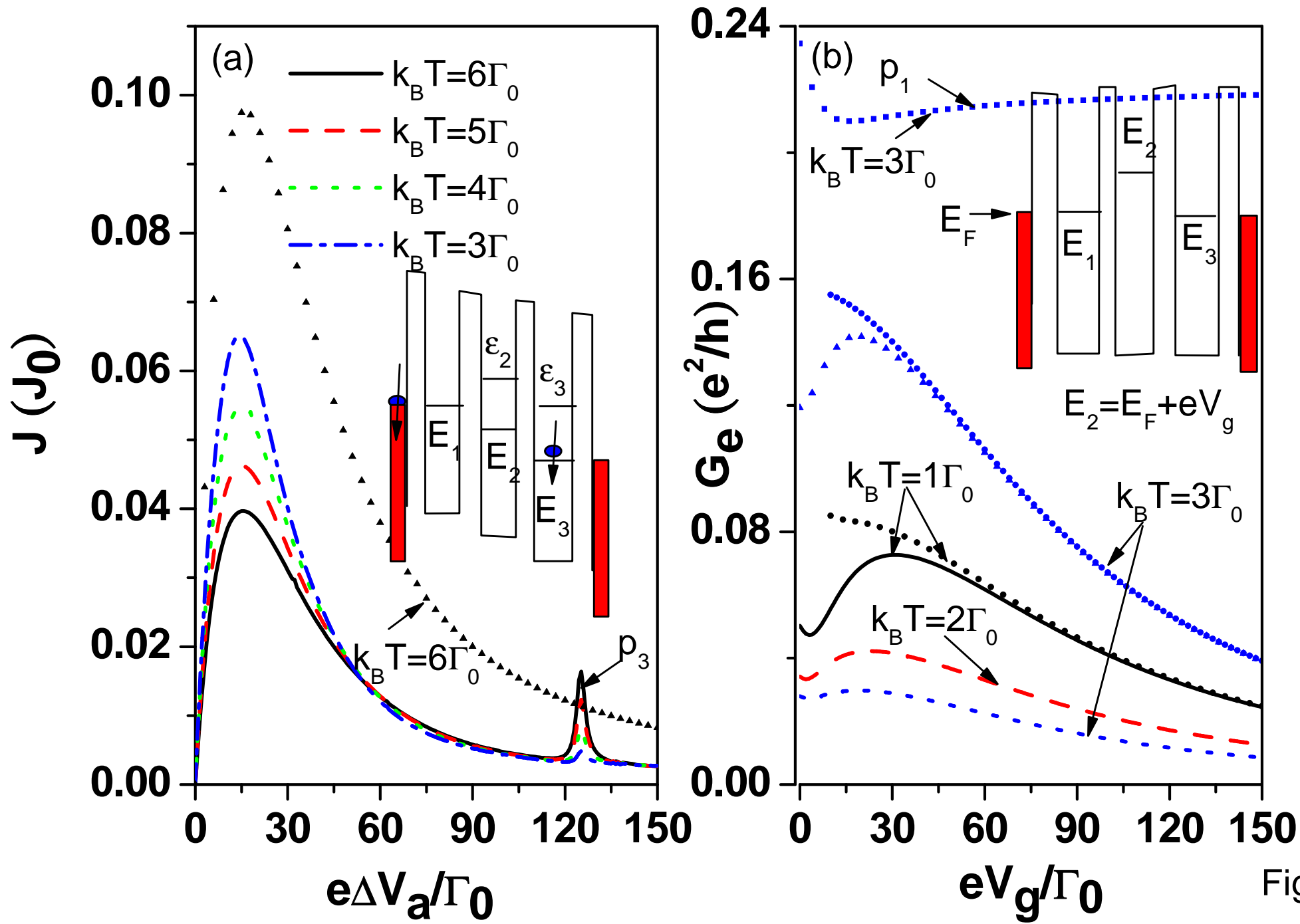
Fig. 3. (a) Tunneling current as a function of applied bias $\Delta
V_a$ for the variation of temperatures, and (b) electrical
conductance ($G_e$) as a function of $E_2-E_F=eV_g$ for various
temperatures with $E_1=E_3=E_F$. In diagram (a) we have
$\epsilon_2=E_2+U_{23}$, $\epsilon_3=E_3+U_{23}$, and
$J_0=2e\Gamma_0/h$.

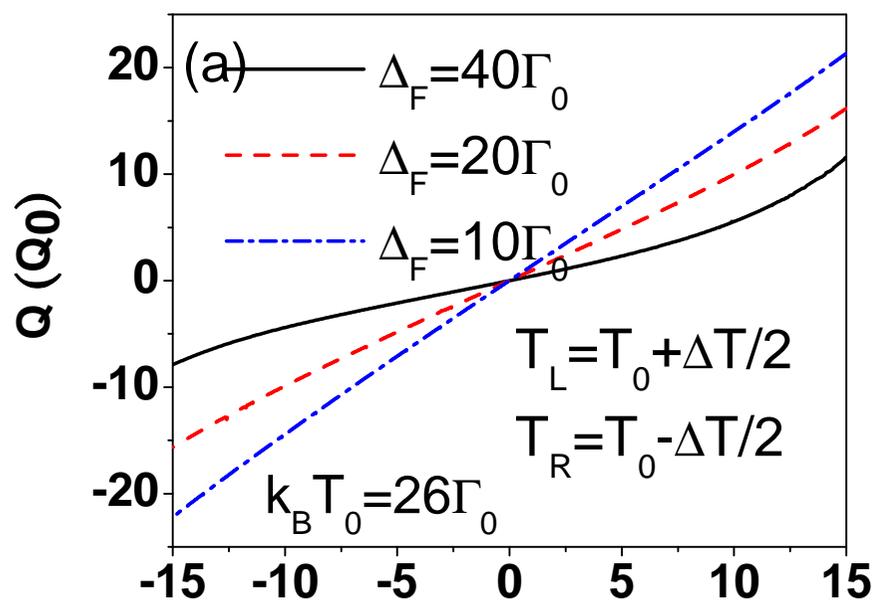
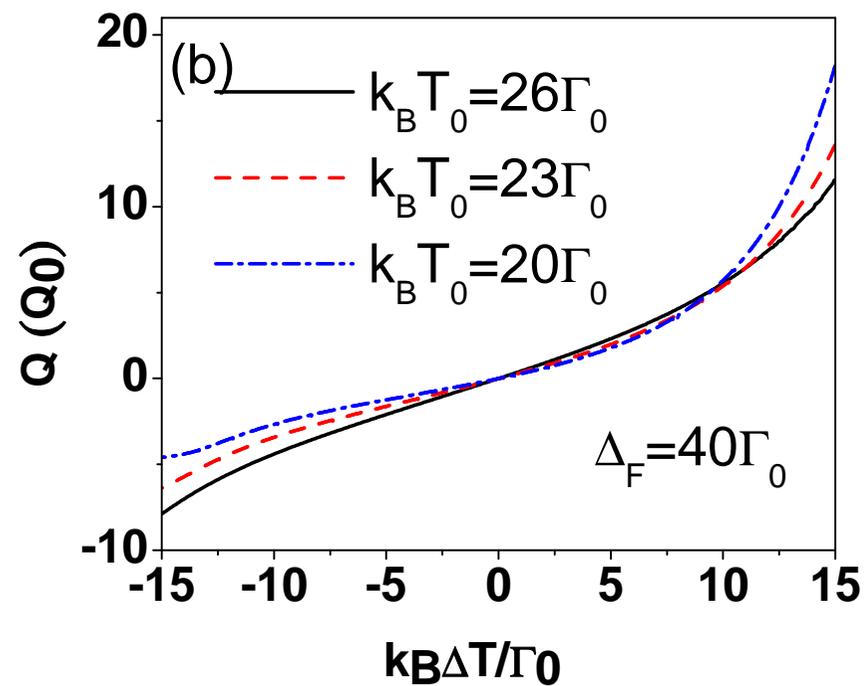
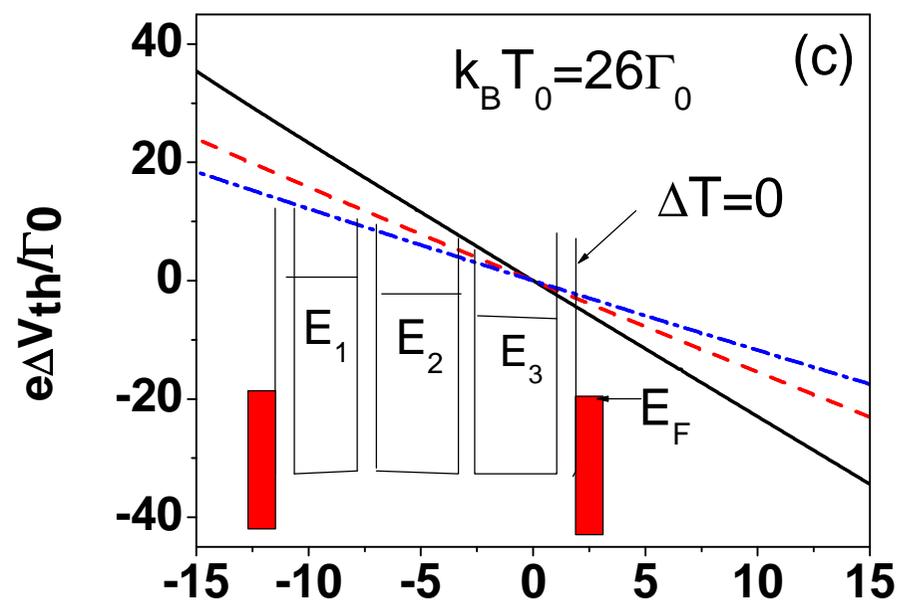
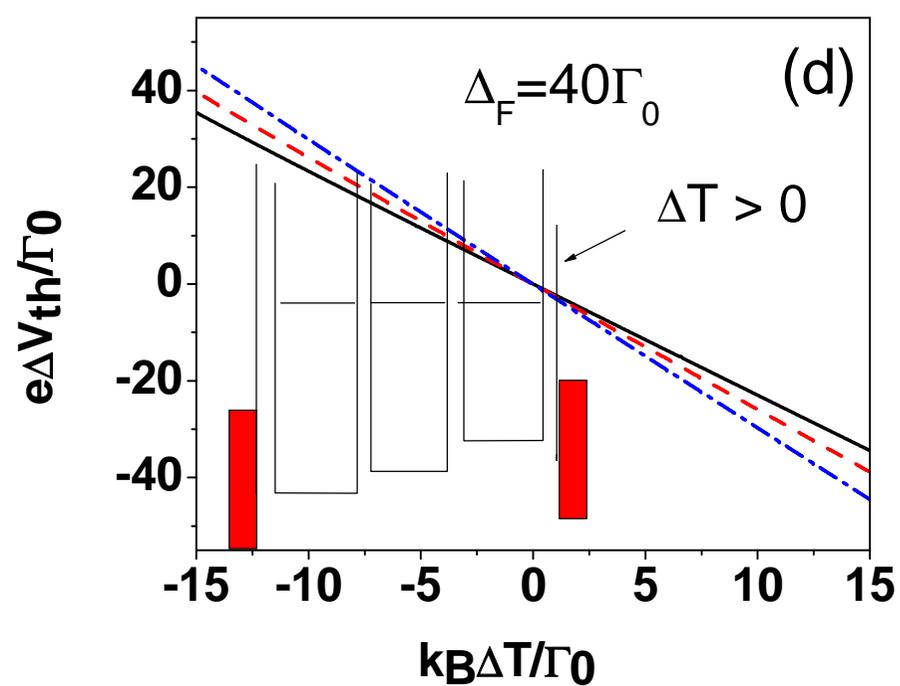
Fig. 4. (a) Electron heat current ($Q=Q_R$) as a function of
temperature bias $\Delta T$ for TQD junction with staircase energy
levels at $k_BT_0=26\Gamma_0$ and $D=8~nm$. (b) $Q$ for different
$T_0$ values. Other physical parameters are the same as those for
the solid line in (a). (c) and (d) are the thermal voltage ($\Delta
V_{th}$) corresponding to (a) and (b), respectively.
$Q_0=\Gamma^2_0/h$.

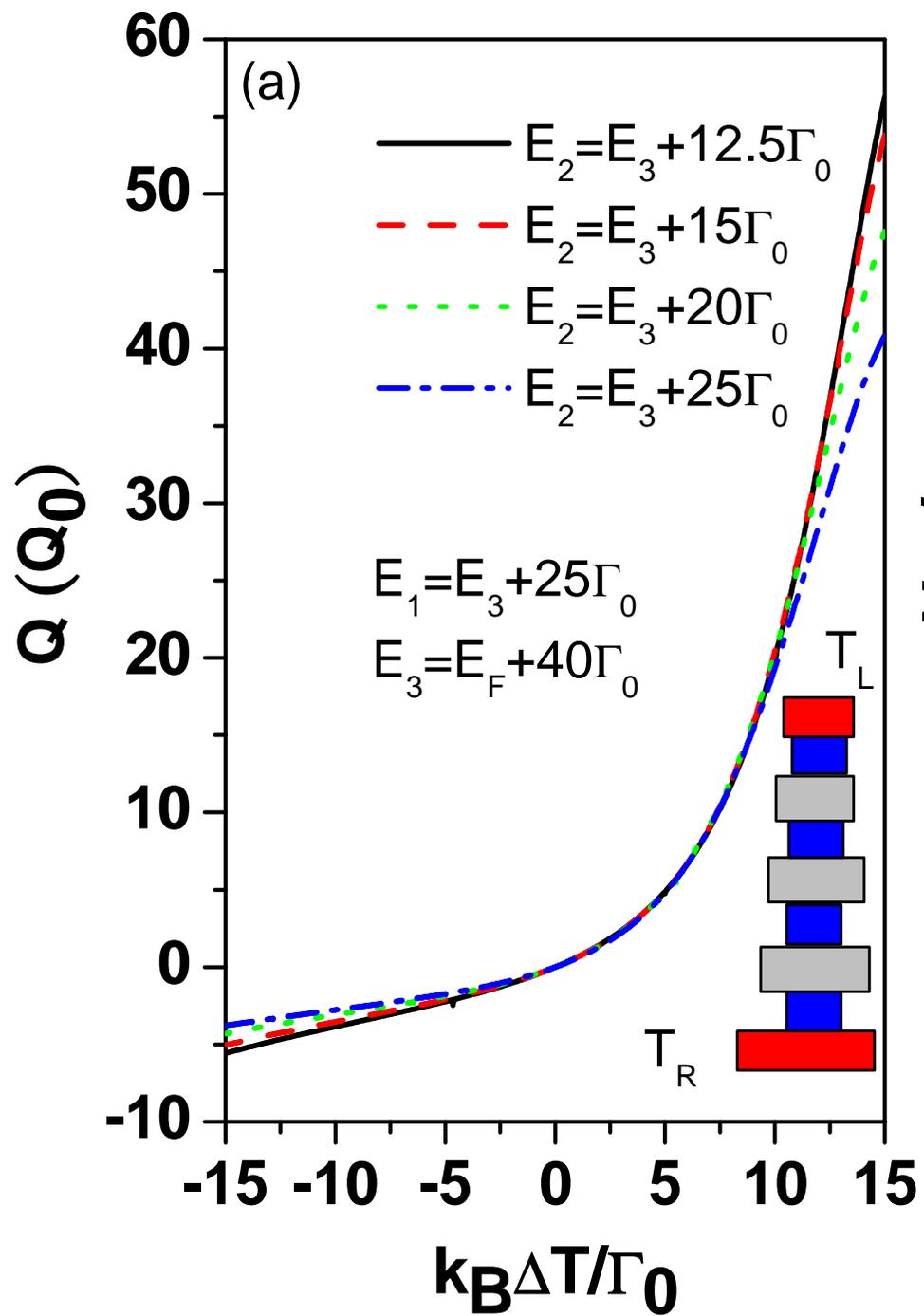
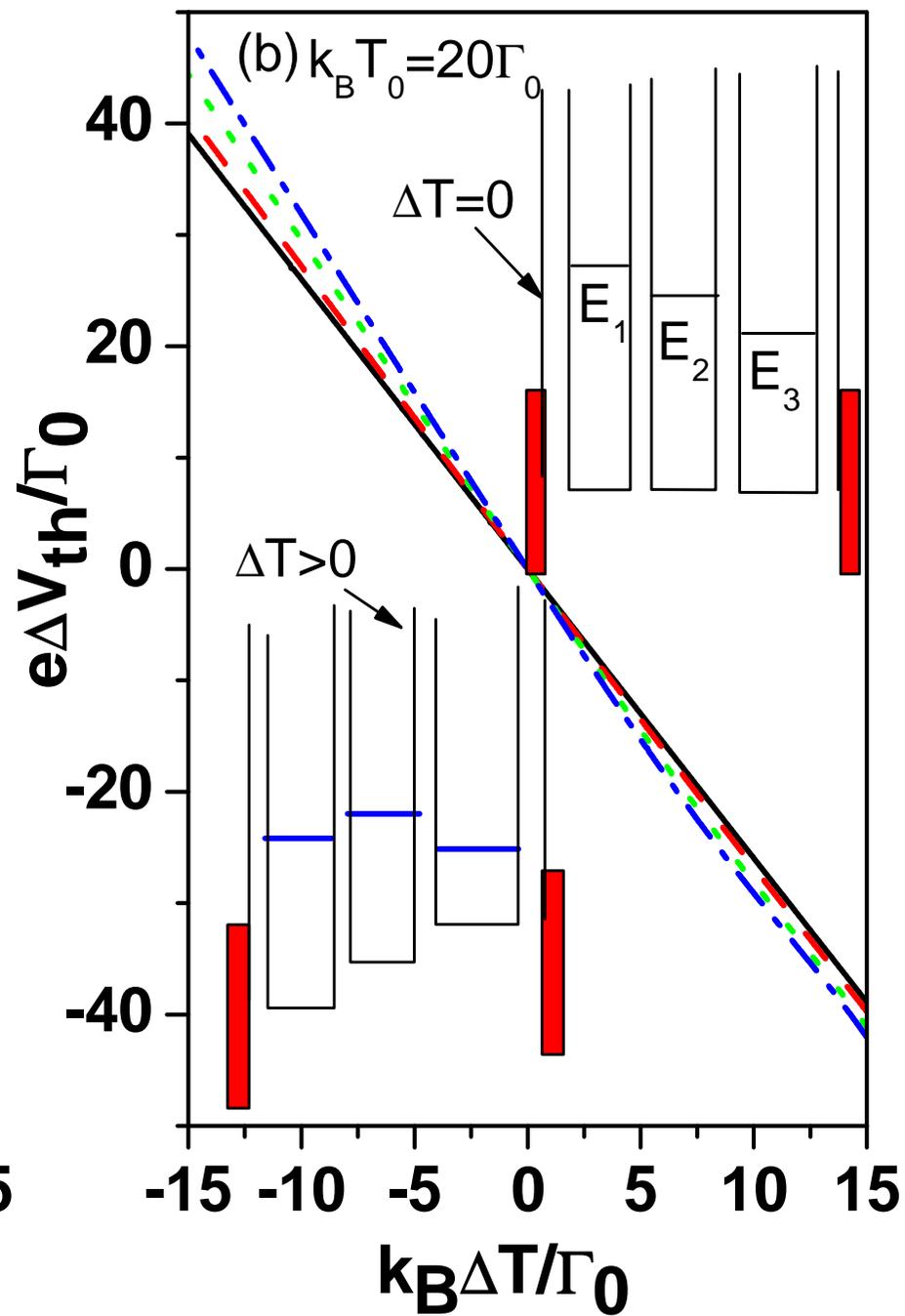
Fig. 5: (a) Heat current (Q), and (b) thermal voltage ($e\Delta
V_{th}$) as a function of temperature bias for different values of
$E_2$ at $E_1=E_3+25\Gamma_0$ and $E_3=E_F+40\Gamma_0$. Other
physical parameters are the same as those of dash-dotted line in
Fig. 4(b).

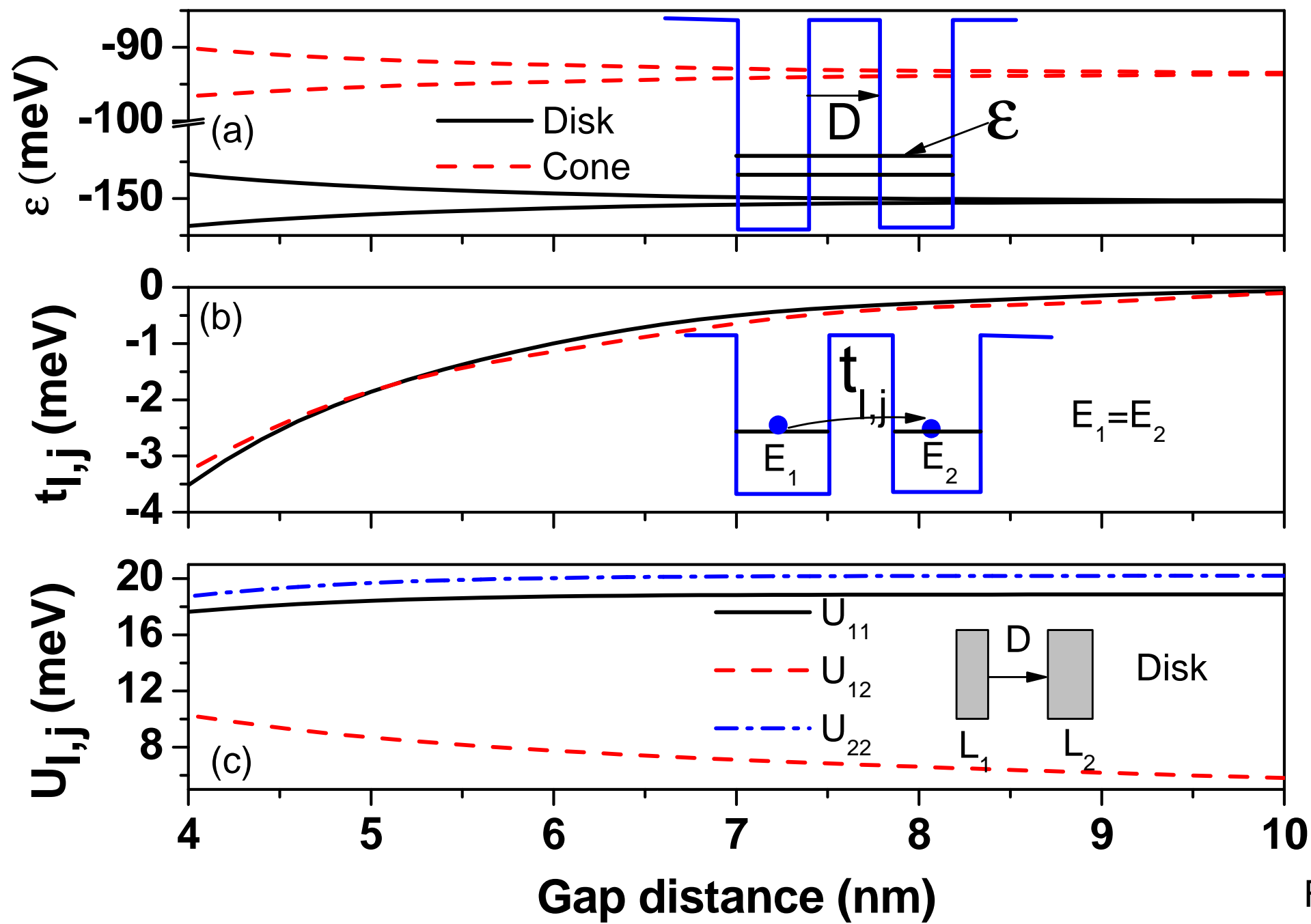
\end{document}